\documentstyle[preprint,aps]{revtex}
\draft
\begin{document}

\title{Pion flow and antiflow in relativistic heavy-ion collisions}
\bigskip
\author{Bao-An Li and C. M. Ko }
\address{Cyclotron Institute and Physics Department,\\
Texas A\&M University, College Station, TX 77843}
\maketitle

\begin{quote}
Within the framework of a relativistic transport model (ART 1.0) for
heavy-ion collisions at AGS energies, we study the transverse flow
of pions with respect to that of nucleons using two complementary approaches.
It is found that in central collisions pions develop a weak
flow as a result of the flow of baryon resonances from which they are
produced. On the other hand, they have a weak antiflow in peripheral
collisions due to the shadowing of spectators. Furthermore,
it is shown that both pion flow and antiflow are dominated by those
with large transverse momenta.
\end{quote}
\newpage
A major highlight of recent experiments at Brookhaven's AGS is the discovery of
sideward collective flow of nucleons in reactions of Au+Au
at $p_{beam}/A$=10.8 GeV/c\cite{e877,zhang}. In view of the large
number of pions created in these reactions and the strong coupling between the
nucleon and pion, it is interesting to know if pions also have a collective
flow behaviour and, if yes, how the pion flow
is related to the nucleon flow. These questions have recently been
addressed in experiments by the E877 collaboration by studying
the correlation between pion transverse momentum distributions in
the reaction plane and the direction of nucleon collective flow near the
projectile rapidity. In these experiments,
the strength and direction of pion collective flow
were measured by examining the multiplicity ratio $R$ of pions emitted in
the direction of nucleon flow over those emitted in the opposite
direction of nucleon flow. The pion flow or antiflow is then
characterized quantitatively by the ratio $R$
being larger or smaller than one.
Surprisingly, from the 13\% most central collisions where a large strength of
nucleon flow was found\cite{e877a}, preliminary data with
relatively large errors show neither flow nor antiflow signatures for pions.
On the other hand, theoretical calculations
using RQMD\cite{rqmd} and ARC\cite{arc} have predicted the existence of
pion antiflow for the minimum biased events in Au+Au reactions.
However, these calculations do not provide the much needed information
about the impact parameter dependence of the pion flow or antiflow in order
to understand the experimental results. In this paper we carry out a
detailed study on the pion flow and antiflow for both central and
peripheral collisions using two complementary approaches. We find
that pions show a weak flow behaviour in central collisions
due to the flow of baryon resonances from which they are produced but
a weak antiflow behaviour in peripheral collisions as a result of the
shadowing of spectators. Moreover, it is shown that both pion
flow and antiflow are dominated by those with large transverse momenta.

Our study is based on the relativistic transport model (ART 1.0)
developed recently for heavy-ion
collisions at AGS energies\cite{art2}.
In this model the phase space distribution functions of
baryons ($N,\Delta(1232),N^{*}(1440),N^{*}(1535),\Lambda,\Sigma$)
and mesons ($\pi,\rho,\omega,\eta,K$) are evolved under the influence of
hadron-hadron scatterings and also an optional mean field for baryons.
The model has been rather successful in studying many aspects of
heavy-ion collisions at AGS energies. We refer the reader to our previous
work for more details \cite{art2}.

In order to identify signatures of pion flow or antiflow,  we use
two complementary approaches. One is based on the analysis of the average
transverse momentum in the reaction plane as a function of rapidity
\cite{dani}, and the other is
the analysis of the strength of flow as a function
of transverse momentum at a constant rapidity \cite{e877a,li95}.
First, we show in Fig.\ 1 the average transverse momentum of nucleons and pions
in the reaction plane as a function of rapidity for both
central (with impact parameters of $ b\leq 4 $ fm) and peripheral
(with impact parameters of 8 fm $\leq b \leq 10$ fm) collisions.
It is seen that the average transverse momentum of pions is much
smaller than that of nucleons. This seems in agreement with
the experimental observations. However, more detailed study reveals that
pions near target or projectile rapidities have a weak flow
behaviour in central collisions and an antiflow behaviour in
peripheral collisions with respect to the flow direction of nucleons.
The reason that  pions flow (antiflow) in central (peripheral) collisions
is due to the fact that the direction of pion flow
(or the sign of average transverse momenta ) is a result of
the competition between the collective flow of baryon
resonances and the shadowing of spectators through rescatterings and
reabsorptions. In our model pions are produced either directly from
particle-particle collisions or from the decay of resonances. In central
collisions pions are produced throughout the whole reaction volume, and
there is thus little shadowing effect. If the colliding particle pairs or
baryon resonances which produce pions have a large flow velocity, then
the produced pions would also have a certain flow velocity in the same
direction as the nucleon as a result of momentum conservation.
However, because of the production kinematics, the pion flow is
much reduced.  On the other hand,
in peripheral collisions the shadowing effect from the spectators
dominates and therefore results in the apparent antiflow of pions in
the opposite direction of nucleon flow. We find that the transition from
pion flow to antiflow occurs at an impact parameter of about 7 fm for
the Au+Au reaction at $p_{beam}/A=$ 10.8 GeV/c.
It is interesting to mention that the transition from pion
flow to antiflow at an impact parameter of about 3 fm has long been
predicted at BEVALAC and/or SIS/GSI energies by several
groups\cite{li91,bass93,li94,dani95} and has been confirmed
recently by experiments\cite{fourpi,eos}. The smaller
shadowing effects at AGS energies as indicated by the smaller pion
transverse momentum and a larger transition impact parameter
at which it becomes important
is due to the fact that the spectators fly away with very large
longitudinal momenta in these reactions and thus have less effects
on pions from the participant region as compared to heavy ion collisions
at SIS/GSI energies.

The above explanation is more clearly demonstrated in Fig.\ 2  where
contributions to the average transverse momentum from  pions already
existing (free pions) and pions from the decay of baryon resonances
which are still present at 20 fm/c (bound pions) are shown separately.
We would like to stress that the
relative multiplicities of the two kinds of pions change gradually
during the reaction as shown in Fig.\ 3
for the reaction of Au+Au at $p_{beam}/A=10.8$ GeV/c
and impact parameters of 2 fm and 8 fm.  Because of the large pion-nucleon
cross section, pions freeze-out rather late. At t=20 fm/c, the
multiplicity ratio of bound/free pions is about 40/420 and 20/140 at
impact parameters of 2 fm and 8 fm, respectively.
{}From Fig.\ 2, it is seen that bound pions have a typical
$S$-shaped transverse
momentum distribution similar to those for nucleons in both central and
peripheral collisions. The apparent flow behaviour of bound pions is
due to the collective flow of baryon resonances from which they are
produced. However, free pions produced earlier either directly
from particle-particle collisions or from decays of
resonances have generally gone through several annihilation-production
cycles which can destroy their collectivity, and more importantly
they also have more chance to be
rescattered by the spectators. In central collisions free pions show
less collectivity than bound pions, but still flow in the same direction
as nucleons. In peripheral collisions, however, these pions
show a distinct antiflow behaviour due to the shadowing of spectators.
The final pion transverse momentum distribution therefore reflects the
complicated reaction dynamics of pion production, reabsorption,
rescattering as well as the collective flow of baryon resonances.
The study of pion flow or antiflow may therefore
reveal interesting information about the in-medium cross sections of
elementary processes involving pions and baryon resonances.

Since the currently available spectrometers at the AGS do not allow
a complete analysis of the average transverse momentum as a function of
rapidity, a complementary approach, i.e, the analysis of the pion
transverse momentum spectrum in the reaction plane at a fixed rapidity,
has been used\cite{e877a}. In this approach,
the ratio $R(p_t)\equiv (dN^+/dp_t)/(dN^-/dp_t)$,
where $N^+(N^-)$ is the number of particles emitted in the reaction
plane in the same (opposite) direction of nucleon sideward flow, is examined
as a function of transverse momentum near the projectile rapidity.
To increase the statistics of our analysis
particles with azimuthal angles smaller
than $20^0$ with respect to the reaction plane and within the rapidity
range of $|y-2.95|\leq 0.35$ are included.
We note that the projectile rapidity of 3.1
for the Au+Au reaction at $p_{beam}/A$=10.8 GeV/c is in this rapidity bin.
It is clear that the observation of
$R(p_t)>1$ at all transverse momenta is an unambiguous
signature of the existence of sideward flow, and its variation
with the transverse momentum can also give
a detailed measure of the strength of collective flow.
This approach was recently found to provide new information about
the collective flow that is complementary to that obtained from
the analysis of the average
transverse momentum in the reaction plane\cite{li95}.

Fig.\ 4 shows the ratio $R(p_t)$ as a function of $p_t$ for pions from
the central and peripheral collisions of the Au+Au reaction
at $p_{beam}/A=10.8$ GeV/c. The solid lines are drown to guide the eyes.
In central collisions, there is a clear tendency of pion flow in the
same direction as nucleons if their transverse momenta
are larger than 0.2 GeV/c. For large transverse momentum pions in
peripheral collisions, we have
$R(p_t)\leq 1$ indicating that there is a tendency of antiflow.
It is also interesting to note that very low transverse momentum pions
with $p_t\leq 0.1$ GeV/c in peripheral collisions are preferentially
emitted in the direction of nucleon flow. From the lower window of
Fig.\ 2 we see that these pions are mainly from
the decay of baryon resonances after the freeze-out. These resonances
generally have light masses and similar transverse momentum distributions
as nucleons. The decay of these resonances therefore emits preferentially
low transverse momentum pions along the direction of nucleon flow.
Although the impact parameters of central collisions studied here are
close to those of the 13\% most central Au+Au collisions studied by the
E877 collaboration, a direct comparison
of our results with the experimental data
requires a careful simulation
of the experimental filters.
Nevertheless, our study indicates that one needs to be cautious in
searching for pion antiflow in central heavy ion collisions.

In summary, we have studied the transverse flow of pions with respect to
that of nucleons using two complementary approaches.
Within the framework of the relativistic transport model (ART 1.0)
for heavy-ion collisions at AGS energies, pions are found to have a
weak flow behaviour due to the flow of baryon resonances
in central collisions. In peripheral collisions pions have instead a
weak antiflow behaviour as a result of the shadowing of spectators.
Furthermore, it is found that both pion flow and antiflow are dominated by
those with large transverse momenta. These results are expected to
be useful for detailed experimental studies of pion flow and antiflow
in relativistic heavy-ion collisions.

We would like to thank P. Danielewicz and G.Q. Li for helpful discussions.
BAL also acknowledges the kind hospitality
extended to him by the nuclear theory group at
Michigan State University where part of this work was carried out.
This work was supported in part by NSF Grant No. PHY-9212209
and the Welch Foundation Grant No. A-1110.  The support of CMK by a Humboldt
Research Award is also gratefully acknowledged, and he would like to thank
Ulrich Mosel of the University of Giessen for the warm hospitality.

\newpage
\section*{Figure Captions}
\begin{description}

\item{\bf Fig.\ 1}\ \ \
The average transverse momentum of nucleons and pions in
the central (upper window) and peripheral (lower window) collisions
of Au+Au at $p_{beam}/A=10.8$ Gev/c.

\item{\bf Fig.\ 2}\ \ \
The average transverse momentum of bound and free pions in
the central (upper window) and peripheral (lower window) collisions
of Au+Au at $p_{beam}/A=10.8$ Gev/c.

\item{\bf Fig.\ 3}\ \ \
The evolution of pion multiplicities in the reactions
of Au+Au at $p_{beam}/A=10.8$ GeV/c and impact parameters
of 2 fm and 8 fm.

\item{\bf Fig.\ 4}\ \ \
The transverse momentum dependence of the strength $R(p_t)$ of
transverse flow in the reactions of Au+Au at $p_{beam}/A=$10.8 GeV/c
and impact parameters of 2 fm and 8 fm.
\end{description}

\end{document}